\pdfoutput=1
\documentclass[sigplan,10pt,nonacm]{acmart}
\setcopyright{none}
\renewcommand\footnotetextcopyrightpermission[1]{}

\usepackage{booktabs}
\usepackage{tabularx}
\usepackage{array}
\usepackage{graphicx}
\usepackage{subcaption}
\usepackage{multirow}
\usepackage{float}
\usepackage{xcolor}
\usepackage{pgfplots}
\pgfplotsset{compat=1.18}
\usepgfplotslibrary{groupplots}
\usepackage{siunitx}
\usepackage{tikz}
\usetikzlibrary{arrows.meta, positioning, fit, backgrounds}
\usepackage{listings}
\usepackage{etoolbox}
\lstdefinelanguage{MLIR}{
  morekeywords={func, module, return, br, cond_br, call, alloc, dealloc, load, store},
  morecomment=[l]{//},
  morestring=[b]",
  sensitive=true,
}
\makeatletter
\def\lst@makecaption#1#2{%
  \def\@captype{table}%
  \@makecaption{#1}{#2}%
}
\makeatother

\usepackage{stfloats}          %

\newcolumntype{C}[1]{>{\centering\arraybackslash}p{#1}}

\begin{document}

\title{Torch-PIM: Automated Profile-Guided PIM Offloading for PyTorch}

\author{Heeeon Lee}
\affiliation{%
  \institution{Yonsei University}
  \country{Republic of Korea}}
\email{lee.heeeon@yonsei.ac.kr}

\author{Hyunwoo Nam}
\affiliation{%
  \institution{Yonsei University}
  \country{Republic of Korea}}
\email{ae04071@yonsei.ac.kr}

\author{Junyong Heo}
\affiliation{%
  \institution{Yonsei University}
  \country{Republic of Korea}}
\email{junyongheo@yonsei.ac.kr}

\author{Hyunmo Sung}
\affiliation{%
  \institution{Yonsei University}
  \country{Republic of Korea}}
\email{banramlo95@gmail.com}

\author{Jay Hwan Lee}
\affiliation{%
  \institution{Yonsei University}
  \country{Republic of Korea}}
\email{jlee758@yonsei.ac.kr}

\author{Yeonsoo Kim}
\affiliation{%
  \institution{Yonsei University}
  \country{Republic of Korea}}
\email{yeonsoo.kim@yonsei.ac.kr}

\author{Seongho Jeong}
\affiliation{%
  \institution{Yonsei University}  
  \country{Republic of Korea}}
\email{seongho.jeong@yonsei.ac.kr}

\author{Shinhyung Yang}
\affiliation{%
  \institution{Kiel University}
  \city{Kiel}
  \country{Germany}}
\email{shinhyung.yang@email.uni-kiel.de}

\author{Bernd Burgstaller}
\affiliation{%
  \institution{Yonsei University}
  \country{Republic of Korea}}
\email{bburg@yonsei.ac.kr}

\renewcommand{\shortauthors}{Lee et al.}
\begin{abstract}
Modern deep learning (DL) workloads are limited by data movement, and
processing-in-memory~(PIM) targets this bottleneck by placing compute units near
the memory. However, PyTorch and other DL frameworks lack compiler support for making this
decision on the code they lower: existing offloading frameworks target
hand-written C/C++ programs, while those that address DL fix the candidate set
to a list of operator types before lowering.

We present Torch-PIM, a compiler framework that uses profile-guided
optimization~(PGO) to decide host-versus-PIM placement over the loop nests that
progressive lowering materializes. Every parallel loop nest the pipeline emits
enters the candidate space, and each is assessed in two stages: the amount of work
it carries, and its memory boundedness. Every quantity the assessment consumes is
profiled on the host or obtained from the
multi-level intermediate representation (MLIR) of the code.

Across PIM configurations of 32 to 128~cores, Torch-PIM's offloading decisions yield
speedups of up to \qty{8.6}{\times} on tensor operators, \qty{2.9}{\times} on MLP,  \qty{4.4}{\times} on  Attention, \qty{5.1}{\times} on  GPT-J-6B, and \qty{3.6}{\times} on LLaMA-7B over CPU-only execution.
\end{abstract}

\maketitle

\section{Introduction}
Modern DL workloads are increasingly limited by data movement rather than arithmetic throughput~\cite{Gholami:2024:MemoryWall,GmezLuna2023,Ye2024}. PIM targets this bottleneck by integrating compute units directly into or near memory. However, PIM introduces a software challenge: deciding which code regions to offload from the host to PIM~\cite{meta}. Incorrect offloading decisions are costly; naive policies based on last-level cache misses per kilo in\-str\-uc\-tions (MPKI) can incur up to 41.6× slowdowns because the context-switch and data-dependency overhead incurred at each transition between the host and PIM are not accounted for~\cite{wei_pimproff}, and manual code offloading is entirely impractical for programmers~\cite{Khan2020A}. Automated compiler-driven offloading is therefore necessary. 

Prior work has demonstrated the advantage of offloading for PIM-integrated systems. However, designing an effective compiler-driven DL offloading framework for such systems faces three challenges.
First, no framework automates a PIM offloading decision for code that is produced from the DL framework after lowering. Only hand-written C/C++ benchmarks, primarily graph workloads, have been the main target of automatic offloading~\cite{PEI,7920847, Hadidi2017CAIRO, TOM, wei_pimproff, 10546698A3PIM,RDPIM,9502483, ALP, IOTPIM}. Although offloading methods for DL operators have been investigated, they restrict the target operators to predetermined candidates and focus on orchestrating the fixed operators based on their strategies, such as subgraph patterns~\cite{PIMFlow, 10.1145/3622781.3674189}, or optimizing the performance on PIM~\cite{OptiPIM,sun_pimcomp,Chen:2023:SimplePIM}.

Second, the granularity of offloadable regions does not fit the structure of DL workloads. DL workloads are expressed as tensor operators, and a single operator commonly has several phases with different behavior, while a model chains hundreds of such operators~\cite{tensorcompiler}. 
At the operator level, compute- and memory-bound loop nests are grouped into one unit: more than half of the candidate layers in a CNN can be placed neither wholly on the host nor wholly on PIM~\cite{PIMFlow}, and offloading such an operator as a whole can result in worse performance than not offloading it~\cite{IOTPIM,10546698A3PIM}. Offloading at the source loop level is impractical, since a tensor operator in DL frameworks carries no loop for a programmer to annotate or an instrumenter to mark~\cite{9502483,loopim}. The basic block level risks frequent CPU--PIM switches: if consecutive blocks are assigned to different targets, every boundary incurs data-transfer and synchronization cost.

Third, the profiling cost for offloading decisions is expensive. Prior work executes entire regions on PIM to decide the best offloading~\cite{wei_pimproff}; executing compute-bound regions on PIM is redundant and expensive. Another work profiles each input separately, which costs up to \SI{11.6}{\times} the execution time of the workload it optimizes~\cite{IOTPIM}.

We present Torch-PIM, a compiler framework that makes offloading decisions over loop nests produced after lowering from PyTorch~\cite{Paszke:2019:PyTorch}. Torch-PIM lowers a PyTorch model through Torch-MLIR~\cite{torchmlir} to \texttt{linalg}, then to explicit loop nests, and admits each parallel loop nest the pipeline produces as an offloading candidate. Nothing designates these candidates: they are not a list of operator types, not patterns identified in advance, and not regions a programmer annotated. Whatever the lowering emits enters the candidate space. One operator therefore yields several candidates, each judged on its own, so an operator that mixes
memory-bound and compute-bound loop nests is split along that line instead of
being placed as a unit. Torch-PIM makes each decision from host-side information alone. For every loop nest it gathers three quantities: the elapsed time of the CPU-only execution;
the achieved DRAM read bandwidth and the store-bound cycle fraction, derived
from five native hardware performance-monitoring events collected through
PAPI~\cite{jagode2025papi} in a single counter set; and the volume of data a PIM placement
would transfer, obtained from the MLIR. These are weighed against a switch overhead
calibrated once on the target platform. No candidate is ever executed or
simulated on the PIM device.

We evaluate Torch-PIM on 15~PyTorch tensor operators, 2~layers of tensor operators, and 2~real-world benchmarks, LLaMA-7B~\cite{Touvron:2023:Llama2}, and GPT-J-6B~\cite{gpt-j}. Torch-PIM's decisions yield speedups of up to \qty{8.6}{\times} on tensor operators, \qty{4.4}{\times} on layers of tensor operators, \qty{5.1}{\times} on  GPT-J-6B, and \qty{3.6}{\times} on LLaMA-7B over CPU-only execution.

In summary, Torch-PIM makes the following contributions:

\begin{itemize}
  \item \textbf{Automated host-vs-PIM offloading for the code produced from DL framework after lowering.} Torch-PIM lowers PyTorch models through MLIR and makes offloading decisions autonomously over the code it generates. 
  \item \textbf{Compiler-materialized loop nests as the decision unit for offloading.} 
  Torch-PIM identifies offloading candidates at the loop nests that progressive lowering materializes, which do not exist until the compiler produces them and therefore do not require an offloading decision at the abstraction level of the input source code.
  Thus, Torch-PIM eliminates the abstraction mismatch between abstract source-level ML operators and their lowered representations, which
  oftentimes contain memory- and compute-bound loops that require individual offloading decisions.
  \item \textbf{Profile-guided offloading decision with micro-architectural events.}
  Torch-PIM employs profiling information obtained from micro-ar\-chi\-tec\-tu\-ral events of the
  underlying host architecture to identify memory-bound loops: elapsed CPU cycles, DRAM
  read bandwidth, and the fraction of cycles a loop spends bound on stores. These are weighed
  against the overhead of offloading, calibrated once on the target platform.
  
  \item \textbf{An offload rule built entirely from host-side measurements.} 
  Torch-PIM reaches an offloading decision without any measurement of PIM-side behavior.
  \item \textbf{Evaluation on real PyTorch framework.} 
  We conduct an extensive experimental evaluation on code generated from PyTorch framework, consisting of 15 PyTorch tensor operators, 2 layers of tensor operators, and 2~real-world benchmarks comprising LLaMA-7B~\cite{Touvron:2023:Llama2}, GPT-J-6B~\cite{gpt-j}. PIM-offloading of
  Torch-PIM achieves speedups of up to \qty{8.6}{\times} on 32, 64, 128~PIM cores over CPU-only execution.
\end{itemize}

\section{Background}

\subsection{Modern DL Workloads and the Viability of PIM}
\label{sec:bg:pim}

Data movement limits performance across a wide range of workloads.
A top-down analysis of 77K functions drawn from 345 applications
identified 144 functions that account for at least 3\% of total cycles
and spend more than 30\% of their execution bound on memory; among
these, the functions whose bottleneck lies in DRAM bandwidth benefit
most from computation placed near memory~\cite{Oliveira2021DAMOV}. The same
behavior appears in DL inference. A layer-level analysis of
neural network models for edge devices attributes a large fraction of
inference time to data movement~\cite{Boroumand2021}, and roofline
analyses of large language models establish that the autoregressive
decode phase operates at low arithmetic intensity and is limited by
memory bandwidth~\cite{yuan2024llm}.
However, identifying the best execution unit solely from workload names
or the operator identities is difficult due to the different
characteristics of the host architectures.
FFN kernels that are compute-bound on datacenter GPUs move into the
memory-bound region on edge devices, where cache capacity and LPDDR
bandwidth are limited~\cite{llm_gpu_char}.

By addressing the memory bottleneck, PIM has attracted attention in academia,
and has moved from academic proposals into commercial silicon.
HBM2~\cite{Kwon:2021:SamsungHBM, aquabolt_xl_pim} and GDDR6~\cite{skh2} were followed
in 2026 by an LPDDR5X-based device~\cite{samsung2026}.
Standardization is also underway: following the JESD209-6 LPDDR6
standard~\cite{jedec_lpddr6}, a LPDDR6 Processing-in-Memory standard is
in development~\cite{jedec2026_lpddr6_pim}. These devices are designed to
minimize changes to the host memory system, so the same physical memory
is reached both through the host's ordinary access path and through the
PIM compute path.

Execution on PIM differs from execution on the host in three respects.
First, compute units placed inside the banks operate under area and
power budgets and are simpler than host cores. The compute unit of a
commercial PIM device is an in-order 32-bit RISC core running at
\SI{350}{MHz} that requires eleven hardware threads to fill its pipeline. Second,
switching execution between the host and PIM carries a cost. Data to be
offloaded may reside in the host caches and must be flushed at the
switch point, and a transfer in the opposite direction is required when
the host reads the result. Third, the host and PIM cannot write the
same data concurrently during an offloaded region. In the commercial
LPDDR5X PIM device, the host must place the device back in
conventional DRAM mode before reading the
result~\cite{samsung2026}.

\subsection{Torch-MLIR}
\label{sec:bg:torchmlir}

Torch-MLIR~\cite{torchmlir} maps a PyTorch model into MLIR. It traces the
model into a graph of tensor operations and emits the \texttt{torch} dialect,
which mirrors the PyTorch operator set with static types attached. Once the
model is in MLIR, it can be lowered through the dialects MLIR provides and the
passes at each level can be varied, so a compiler for a new target can be built
by choosing where to intervene rather than by writing a backend from scratch.

\section{Related Work}
\subsection{Existing Automatic PIM Offloading Frameworks}
Automatic PIM offloading frameworks differ in the granularity at which they form candidate regions, from individual instructions to entire kernels, and in the information their decision rules consume. In all of them, however, the candidate set is fixed by boundaries of the source program or the framework interface already defines. At the finest granularity, PEI~\cite{PEI} defines a set of memory-side primitives and dispatches individual operations to them according to their locality. GraphPIM~\cite{7920847} maps atomic instructions in graph workloads onto memory-side execution. PIMProf~\cite{wei_pimproff} partitions a program into basic blocks and functions, simulates each on both sides to obtain per-unit costs, and solves for a partition that accounts for the data movement induced at the boundaries. A$^3$PIM~\cite{10546698A3PIM} uses the same granularity as PIMProf, and classifies functions from a static analysis of their access patterns, and RDPIM~\cite{RDPIM} derives partitions for graph applications from the connectivity of the input graph. CoPIM~\cite{9502483} decides over programmer-written loop bodies, sampling a fraction of the iterations at runtime to estimate the cache behavior of the whole; LooPIM~\cite{loopim} and other loop-oriented approaches~\cite{Maity2023,Maity2025} follow a similar path.

The unit these frameworks decide over is, in every case, one whose boundaries the source determines. The compiler recovers instructions and basic blocks, but their number and extent follow from the control flow the programmer wrote; functions and loop bodies are written in the source directly; operators are fixed by the framework interface. No such correspondence holds for a compiled DL model. A tensor operator specifies an iteration space and nothing more; the pipeline's fusion and decomposition decisions determine how many loop nests execute it, in what order, and where one ends and the next begins. The evidence these frameworks collect is likewise unavailable at compile time: a static analysis of access patterns alone does not establish that a region is bound by memory on the machine that runs it~\cite{10546698A3PIM}, sampling a subset of iterations characterizes only the iterations sampled~\cite{9502483}, and simulating both sides of the execution costs more than a compilation pipeline can absorb~\cite{wei_pimproff}.

\subsection{PIM Compilers for DL Frameworks}
Several compiler infrastructures provide a pipeline for DL frameworks targeting PIM, yet they universally assume the offloading decision is already made, and optimize resource distribution on PIM cores. OptiPIM~\cite{OptiPIM} lowers PyTorch through Torch-MLIR down to linalg and affine loop nests. It formulates an integer linear programming (ILP) problem to map tiling, data layouts, and indexing onto PIM resources, supposing that the offloading decision has already been made. Other frameworks make the same assumption. ATiM~\cite{10.1145/3695053.3731096} extends TVM to auto-generate kernels and auto-tune intra- and inter-DPU optimization spaces for UPMEM, but operates on operations already designated for offloading. CINM~\cite{10.1145/3622781.3674189} provides an MLIR dialect layer for compute-in-memory and compute-near-memory devices, but it doesn't consider ``whether to offload''. Likewise, other end-to-end DNN compilers~\cite{sun_pimcomp} and programming frameworks~\cite{Chen:2023:SimplePIM, comprehensive_pim_dnn} automate data distribution and kernel execution (how to offload), yet leave the decision of what to offload entirely to the programmer. Torch-PIM fills this gap by autonomously deciding CPU-PIM placement across the loop nests it lowers from the DL framework.

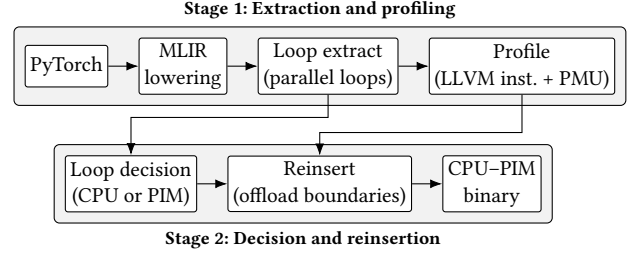
\begin{figure}[t]
\centering
\footnotesize
\begin{tikzpicture}[
    node distance=3mm and 4mm,
    box/.style={
        draw,
        rounded corners=1pt,
        align=center,
        inner sep=2pt,
        minimum height=5.5mm,
        fill=white
    },
    arr/.style={-{Latex[length=1.6mm]}, thin},
    stagebox/.style={
        draw,
        rounded corners=2pt,
        fill=gray!10,
        inner sep=4pt
    }
]

\node[box] (input) {PyTorch};
\node[box, right=of input] (lower) {MLIR\\lowering};
\node[box, right=of lower] (extract) {Loop extract\\(parallel loops)};
\node[box, right=of extract] (profile) {Profile\\(LLVM inst.\ + PMU)};

\node[box, below left=8mm and 8mm of extract] (decision) {Loop decision\\(CPU or PIM)};
\node[box, right=of decision] (reinsert) {Reinsert\\(offload boundaries)};
\node[box, right=of reinsert] (binary) {CPU--PIM\\binary};

\draw[arr] (input.east) -- (lower.west);
\draw[arr] (lower.east) -- (extract.west);
\draw[arr] (extract.east) -- (profile.west);

\draw[arr] (decision.east) -- (reinsert.west);
\draw[arr] (reinsert.east) -- (binary.west);

\draw[arr] (extract.south) -- ++(0,-3mm) -| (decision.north);
\draw[arr] (profile.south) -- ++(0,-5mm) -| (reinsert.north);

\begin{scope}[on background layer]
    \node[stagebox, fit=(input)(lower)(extract)(profile),
          label={[font=\scriptsize\bfseries]north:Stage 1: Extraction and profiling}] {};
    \node[stagebox, fit=(decision)(reinsert)(binary),
          label={[font=\scriptsize\bfseries]south:Stage 2: Decision and reinsertion}] {};
\end{scope}

\end{tikzpicture}
\caption{Torch-PIM overview. Stage~1 lowers PyTorch code to MLIR, extracts parallel loop candidates, and profiles them using LLVM instrumentation and PMU events. Stage~2 assigns loop-level CPU/PIM decisions and reinserts offload boundaries into code generation.}
\label{fig:torchpim-overview}
\end{figure}

\section{Torch-PIM Design}

\subsection{Overview of Torch-PIM}
Torch-PIM is a static-shape DL compiler for improving performance on CPU-PIM integrated systems. Torch-PIM operates in two stages. The first stage lowers operations and characterizes each candidate loop nest both statically and dynamically: static analysis at the MLIR level, and profiling at the LLVM IR level with hardware performance counters. The second stage determines the CPU--PIM execution boundary: Torch-PIM applies the offloading rule to the collected characteristics and emits offload markers in the generated program.

\subsection{Progressive Lowering in MLIR}
Torch-PIM progressively lowers a PyTorch model through MLIR dialects,
identifying candidate loop nests and their static characteristics along the way.
The remaining inputs are measured by profiling the instrumented program on the host.
\label{sec:bg:mlir}

\paragraph{\texttt{torch}: obtains input shapes.}
A PyTorch model is first lowered to the \texttt{torch} dialect by Torch-MLIR~\cite{torchmlir}. To do so, Torch-PIM requires every tensor extent to be a compile-time constant, which
holds when a model is exported for a fixed input shape. The same constant later
appears as a dimension of a \texttt{memref} type, as a loop bound materialized
from it, and as an allocation size in the emitted code.

\begin{figure}[H]
\begin{lstlisting}[language=MLIR,
  basicstyle=\footnotesize\ttfamily,
  columns=fullflexible, keepspaces=true,
  xleftmargin=1.5em, numbers=left, numberstyle=\tiny\color{gray}]
#map = affine_map<(d0, d1) -> (d0, d1)>
#par = ["parallel", "parallel"]

       {indexing_maps = [#map, #map],
        iterator_types = #par}
       ins (%
       outs(%
  ^bb0(%
    linalg.yield %
} -> tensor<64x64xf32>
\end{lstlisting}
\caption{One loop nest at the \texttt{linalg} dialect. The nest states its read
set in \texttt{ins} and its write set in \texttt{outs}, and both operands carry a
constant shape, so the bytes a placement would have to move follow from the
operand types alone: $64\times64\times4 = 16{,}384$ on each side.}
\label{fig:linalg-nest}
\end{figure}

\paragraph{\texttt{linalg}: extracts loop-nest boundaries, logical oper\-ands.}
Torch-PIM first identifies loop nests at the \texttt{linalg} dialect, where
the boundaries of its decision units are settled. Fusion and decomposition decisions act on
structured operations, so how many loop nests are extracted and where their
boundaries fall is determined at this level: a \texttt{softmax} becomes eight loop nests and
an \texttt{attention} layer fourteen. No function of the input program corresponds to any of
them.

Torch-PIM also determines at this level how many bytes a loop nest would have to move
if it were placed on PIM. Each of these nests appear here as a single \texttt{linalg}
operation that declares what it reads and what it writes as whole tensors, through its
\texttt{ins} and \texttt{outs} operands like the example depicted in Fig.~\ref{fig:linalg-nest}. The read and write set of a nest are
therefore available as logical units with a static extent, before lowering scatters
them into individual memory accesses. The size of each operand follows from the product
of its constant shape dimensions times the element width. Torch-PIM then bufferizes
these operations.

Immediately after bufferization, once \texttt{memref} types are settled and before
following passes erase them, Torch-PIM records two things for each operand: whether it is
a \texttt{memref}, and the allocation it resolves to. Operands that are not
\texttt{memref} values are charged no transfer.
Neither is available before this point, because \texttt{linalg} operands are values,
not storage. The
\texttt{ins} and \texttt{outs} of a nest name the tensors it reads and writes, from which the sizes are known, but a tensor at this level has no address: two
operands holding the same tensor may end up sharing one buffer or occupying two, and
an operation that appears to produce a new tensor may in fact overwrite its input.
Bufferization decides the placement, assigning a buffer only to the values that need
one and leaving the rest in place. It settles which data would
actually cross a host--PIM boundary. Lowering to the LLVM dialect afterward replaces
every \texttt{memref} operand with a bare pointer and address arithmetic, and the
information disappears.

Several operands can name the same buffer. An
output written in place carries the buffer of the input it overwrites, and a value
threaded through successive operations reappears as an operand of each. Identifying
operands by their allocation, a buffer that is read and then written in place counts
as one transfer instead of two, and a buffer that a producer writes and a consumer within
the same segment reads back is counted once.

\paragraph{\texttt{scf}: counts iteration.}
Torch-PIM reads the \texttt{scf} dialect and counts iterations of each loop nest.
\texttt{linalg} declares which dimensions of a nest are parallel and which are
reductions, but that is the dependence structure of the iteration space, not its
extent. Bufferization materializes that space into actual loops: parallel dimensions
become \texttt{scf.parallel}, reduction dimensions \texttt{scf.for} with
\texttt{scf.reduce}. The two kinds of iteration are now distinct operations, each
carrying constant bounds.

Only the loops materialized as \texttt{scf.parallel} are passed to the offloading rule,
and the bounds and stride of each give its iteration count,
e.g., \texttt{0 to 64
step 1} yields 64. From these, Torch-PIM collects three values: the product of the
iteration counts over the parallel dimensions alone, the product over every dimension
of the nest, and the trip count of the enclosing loops. The first two feed the
stage-one conditions of Section~\ref{sec:design:gate1}; the third, together with the
buffer sizes of the previous step, feeds the data-movement calculation of
Section~\ref{sec:design:data_move_cost}.

\lstdefinelanguage{LLVM}{
  morekeywords={call,void,ptr,i32,i64,define,declare,getelementptr,store,load},
  morecomment=[l]{;},
  sensitive=true,
}
\lstdefinestyle{llvmir}{
  language=LLVM,
  basicstyle=\ttfamily\scriptsize,
  keywordstyle=\bfseries,
  commentstyle=\itshape\color{gray},
  columns=fullflexible,
  keepspaces=true,
  breaklines=true,
  breakatwhitespace=true,
  breakindent=1.5em,
  postbreak=\mbox{\textcolor{gray}{$\hookrightarrow$}\space},
  xleftmargin=0.5em,
  aboveskip=0.6em,
  belowskip=0.6em,
  frame=tb,
  framerule=0.4pt,
  framesep=4pt,
}

\subsection{Profile-Guided Optimization}
\label{sec:PGO}

\begin{table}[t]
\caption{Native events programmed as a single set on the profiling host.}
\label{tab:pmu}
\footnotesize
\begin{tabular}{@{}ll@{}}
\toprule
Event & Term \\
\midrule
\texttt{offcore\_requests.all\_data\_rd} & off-core data reads \\
\texttt{EXE\_ACTIVITY.BOUND\_ON\_STORES} & store-bound cycles \\
\texttt{CYCLE\_ACTIVITY.STALLS\_TOTAL}   & backend-bound term \\
\texttt{EXE\_ACTIVITY.1\_PORTS\_UTIL}    & backend-bound term \\
\texttt{EXE\_ACTIVITY.2\_PORTS\_UTIL}    & backend-bound term \\
\bottomrule
\end{tabular}
\end{table}

Torch-PIM obtains the inputs to its offloading decision by instrumenting the lowered LLVM~IR with marker calls around each candidate loop nest and profiling it on the host. The markers are thin wrappers built on PAPI~\cite{jagode2025papi}. At the entry marker, the library reads the counters of
a single event set; at the exit marker, it reads them again and accumulates the
difference. 

Torch-PIM collects the following information during profiling. The first is the elapsed time that a region accumulates over
all of its invocations, taken from a timing run that executes the nests with the counters disabled. The other two come from the counter
run, whose events are listed in Table~\ref{tab:pmu}. Off-core data reads give the traffic the region issues past the last-level cache. The remaining four are the counters that the store-bound of the Top-down Microarchitecture Analysis
(TMA)~\cite{intel_topdown} is computed from. No counter is multiplexed and no record is scaled up from a sampled fraction of the execution.

\subsection{Offload Decision Rule}
Torch-PIM decides placement in two stages. The first excludes regions that
cannot recover the minimum offloading costs, or whose iterations cannot be
distributed across the PIM cores; the second selects those that are memory-bound on the host.
\begin{equation}
\label{eq:rule}
\mathrm{offload}(r) \;=\; \mathrm{g\_work}(r) \;\wedge\; \mathrm{g\_mem}(r)
\end{equation}

\paragraph{Stage 1: Minimum Work}
\label{sec:design:gate1}

A region $r$ passes the first stage only if it satisfies both of the following:
\begin{align}
\label{eq:gwork}
\mathrm{g\_work}(r) \;&=\; \text{(C1)} \;\wedge\; \text{(C2)} \\
\label{eq:c1}
\text{(C1)}\quad & T_\mathrm{cpu}(r) \;\ge\; T_\mathrm{ctx} + T_\mathrm{pim}^{\min}(r) \\
\label{eq:c2}
\text{(C2)}\quad & N_\mathrm{iter}(r) \;\ge\; N_\mathrm{core}
\end{align}
A region that fails either condition is placed on the host without evaluating
the second stage.

For offloading to pay off, the host execution time must exceed the sum of two
expenditures: the cost of switching execution to the memory side,
$T_\mathrm{ctx}$; the cost of moving the operands; and the time the kernel takes
on the PIM cores, $T_\mathrm{pim}(r)$. C1 charges the first and the third. The
third cannot be known without executing the region, but it can be bounded from
below.

A loop body that executes on PIM contains at least one instruction, so
$N_\mathrm{total}(r)$ iterations issue at least $N_\mathrm{total}(r)\cdot
I_\mathrm{min}$ instructions; we take $I_\mathrm{min}=1$. The target PIM core is
single-issue and in-order, so at most one instruction retires per cycle, and
distributing the iterations perfectly across $N_\mathrm{core}$ cores leaves an
execution time of at least $N_\mathrm{total}(r)\cdot I_\mathrm{min} /
(N_\mathrm{core}\cdot f_\mathrm{pim})$. The quantity sets memory access latency,
bank conflicts, and load imbalance to zero, so no execution can fall below it.
The threshold in C1 is therefore a lower bound on the true threshold, and a
region that fails C1 cannot satisfy the true condition either.

\paragraph{Stage 2: Memory Boundedness}
\label{sec:design:gate2}
The second stage confirms memory boundedness of each loop from the counter run of Section~\ref{sec:PGO}
\begin{align}
\label{eq:gmem}
\mathrm{g\_mem}(r) \;&=\; \text{(C3)} \;\vee\; \text{(C4)} \\
\label{eq:gbw}
\text{(C3)}\quad & BW_\mathrm{read}(r) =
  \frac{n_\mathrm{rd}(r)\cdot L}{T_\mathrm{cpu}(r)} \;>\; \theta \\
\label{eq:gstore}
\text{(C4)}\quad & \mathrm{StoreBound}(r) =
  \frac{c_\mathrm{store}(r)}{c_\mathrm{backend}(r)} \;>\; \sigma
\end{align}
Here $n_\mathrm{rd}$ is the number of off-core data read requests, $L = 64$\,B is
the cache line size, and $c_\mathrm{store}$ is the cycles bound on stores; the
backend term is the sum of the three backend-bound events and $c_\mathrm{store}$,
following the definition in the Top-down Microarchitecture Analysis
(TMA)~\cite{intel_topdown}. The native events are those in
Table~\ref{tab:pmu}.

The two paths are separate because reads and writes appear in different counters.
Read traffic is recorded directly as off-core read requests. These are the
requests that leave the last-level cache, so multiplying by the cache line size
and dividing by the elapsed time gives the DRAM read bandwidth the region
actually achieved.

Non-temporal stores bypass the cache and so never appear in the
off-core read counter; a region dominated by writes shows $BW_\mathrm{read}
\approx 0$ and cannot pass \eqref{eq:gbw}. The write side is therefore detected
not by traffic volume but by the stall it produces, namely the fraction of
backend cycles bound on stores. Each path covers regions the other cannot
observe, and using only one of them leaves the other side entirely undetected.
We set the two thresholds as follows. We sweep $\theta$ over 0.50--4.00\,GB/s
on the 44~profiled loops, of which 29 benefit from offloading under end-to-end
CPU/PIM execution. Below 1.73\,GB/s, compute-bound regions pass, 
and false positives rise to 3--4; at and above 1.80\,GB/s they reach their
minimum of~2. In the other direction, false negatives grow from 6 to 15 at
2.25\,GB/s and accuracy falls to~0.614. We therefore adopt $\theta =
2.0$\,GB/s, the midpoint of the $[1.80, 2.13]$\,GB/s interval over which false
positives are minimal, which is $0.12 \times B_\mathrm{cpu}$ for the
$B_\mathrm{cpu} = 16.754$\,GB/s of our platform and is carried to other
platforms as that ratio. At this setting, 26 regions are selected, and the
decision matches end-to-end execution for $37/44 = 0.841$ of them. For
$\sigma$ we take~0.4, the most conservative value of the range at which TMA
classifies a region as store bound~\cite{intel_topdown}.

\section{Evaluation}
\subsection{Experimental Setup}
\label{sec:eval:setup}
\begin{table}[t]
\centering
\small
\caption{System configuration.}
\label{tab:tab_system_config}
\begin{tabular}{@{}ll@{}}
\toprule
\multicolumn{2}{@{}c@{}}{\textbf{Out of order CPU}} \\
\midrule
Processor & Intel Xeon Gold 6330 (Ice Lake-SP) \\
Cores     & 2 sockets $\times$ 28 cores, 2 threads/core \\
Clock     & 2.00\,GHz base, 3.10\,GHz max \\
Caches    & 48\,kB L1D, 32\,kB L1I, 1.25\,MB L2 per core; \\
          & 42\,MB L3 per socket \\
\midrule
\multicolumn{2}{@{}c@{}}{\textbf{General-purpose in-order PIM cores}} \\
\midrule
Cores     & 32, 64, 128 \\
Clock     & 1\,GHz, single-issue \\
Caches    & 32\,kB L1I, 32\,kB L1D \\
\bottomrule
\end{tabular}
\end{table}

\definecolor{cTensorOp}{HTML}{DCDCF0}
\definecolor{cLayer}{HTML}{D7EBDC}
\definecolor{cReal}{HTML}{F7E0CC}

\newcolumntype{L}[1]{>{\raggedright\arraybackslash}p{#1}}
\newcommand{\catbox}[1]{%
  \tikz[baseline=-0.1ex]\fill[#1,draw=black!40](0,0)rectangle(1.6ex,1.6ex);}

\begin{table}[t]
\centering
\caption{Benchmark categories in our evaluation. Category labels (a)--(f)
correspond to the panels of Figure~\ref{fig:grid}.}
\label{tab:tab_benchmark_summary}
\small
\renewcommand{\arraystretch}{1.15}
\setlength{\tabcolsep}{4pt}
\begin{tabular}{@{}L{1.95cm}L{2.35cm}p{3.1cm}@{}}
\toprule
\textbf{Category} & \textbf{Benchmarks} & \textbf{Explanation} \\
\midrule
\multicolumn{3}{@{}l}{\catbox{cTensorOp}\;\textbf{Tensor operator}}\\
\addlinespace[2pt]
Elementwise / streaming
& \texttt{ew\_add}, \texttt{copying}, \texttt{scale}, \texttt{zeroing}, \texttt{relu}
& Streaming kernels with no reuse; the case offloading should always take. \\
\addlinespace[2pt]
Transcendental
& \texttt{silu}, \texttt{gelu}, \texttt{tanh}
& Elementwise shape but compute-bound bodies; the case offloading must reject. \\
\addlinespace[2pt]
Reduction / normalization
& \texttt{softmax}, \texttt{layernorm}, \texttt{rmsnorm}
& Operators that lower to loop nests of mixed boundedness, requiring per-nest
  decisions. \\
\addlinespace[2pt]
GEMM / conv primitive
& \texttt{mac}, \texttt{gemv}, \texttt{matmul}, \texttt{depthwise conv}
& Reuse-bearing kernels spanning the decision boundary as $N_\mathrm{core}$
  varies. \\
\midrule
\multicolumn{3}{@{}l}{\catbox{cLayer}\;\textbf{Layer of tensor operators}}\\
\addlinespace[2pt]
Composite layer
& \texttt{attention}, \texttt{MLP}
& Multi-operator layers used to evaluate selectivity beyond single operators. \\
\midrule
\multicolumn{3}{@{}l}{\catbox{cReal}\;\textbf{Real-world workload}}\\
\addlinespace[2pt]
Real LLM decode layer
& \texttt{GPT-J-6B}, \texttt{LLaMA-7B}
& Decode-layer inference \\
\bottomrule
\end{tabular}
\end{table}

\paragraph{Benchmarks and comparisons}
We evaluated Torch-PIM on 15 tensor operators and two layers in PyTorch~\cite{Paszke:2019:PyTorch} and two real-world DL models: GPT-J~\cite{gpt-j} and LLaMA~\cite{Touvron:2023:Llama2}.
We compare our method with three different offloading configurations, as explained below.
\begin{itemize}
    \item \textbf{CPU-only execution (baseline)}: no loop is offloaded, and the entire workload remains CPU-resident.
    \item \textbf{Basic-block-level offloading}: the same decision rule is applied to basic blocks, so each block is judged independently.
    \item \textbf{Operator-level offloading}: the same decision rule is applied at the granularity of the operators the model is written in, so that all loop nests in a single operator are governed by the same decision.
    \item \textbf{Torch-PIM}: Our loop-level offloading method.
\end{itemize}
The three offloading configurations differ only in the unit at which the decision
is made.

\paragraph{System}
We target PIM architectures that place general-purpose programmable cores in memory.
Table~\ref{tab:tab_system_config} summarizes the heterogeneous system configuration used in the simulation. The CPU side is a real Ice Lake machine, a high-performance
server processor, and all host-side times are measured on it bare-metal. The PIM performance is measured on the PIMProf~\cite{wei_pimproff} simulator, which is on the Sniper simulator~\cite{sniper}. Each side of the
heterogeneous system is measured by the means that is most reliable for it.
Regions that remain on the host are measured on real hardware, where cache
behavior, prefetching, and memory-level parallelism are those of a real machine
rather than a model of one. Regions placed on the PIM have no
corresponding hardware and are therefore measured under simulation.
We sweep the number of cores in PIM from 32 to 128 in evaluating workloads
to identify the impact of parallelism in PIM on performance improvement.

\paragraph{Data-movement cost.}
\label{sec:design:data_move_cost}
Torch-PIM assembles the data-mo\-ve\-me\-nt cost of a placement from what the lowering
pipeline collected.
Offloading a loop nest to PIM costs three things: the transitions between host and
PIM, the bytes of the nest's \texttt{ins} that must cross to PIM before it runs there,
and the bytes of its \texttt{outs} that must cross back afterward. Writing $N_{\mathrm{region}}$ for the number of host--PIM transitions
the placement induces, $\mathit{fetch}$ and
$\mathit{flush}$ for those two sets of bytes, and $B_{\mathit{rd}}$ and
$B_{\mathit{wb}}$ for the effective bandwidths at which they move, that cost is
\begin{equation}
T_{\mathrm{copy}} = \mathit{fetch}/B_{\mathit{rd}} + \mathit{flush}/B_{\mathit{wb}} .
\label{eq:tcopy}
\end{equation}
The per-allocation sizes and trip counts of Section~\ref{sec:bg:mlir} give
$\mathit{fetch}$ and $\mathit{flush}$; the remaining terms do not come from the IR.
$N_{\mathrm{region}}$ is an output of the decision rule, since it depends on which
regions the rule places on PIM and where the boundaries between them fall. The PIM
cores share DRAM with the host, so the movement $B_{\mathit{rd}}$ and
$B_{\mathit{wb}}$ describe traffic between the host cache hierarchy and PIM.
Bank-level aggregate bandwidth on the PIM side is reported at the
terabyte-per-second scale on real systems~\cite{PrIM}, more than an order of magnitude
above host channel bandwidth. We therefore use the bandwidth of the host memory system
for both, fixed once per target platform.

\emph{This is an upper bound.} Equation~\ref{eq:tcopy} does not estimate the movement
cost but bounds it from above, for two reasons. First, it charges the entire read set
of every region placed on PIM to $\mathit{fetch}$ and its entire write set to
$\mathit{flush}$. It does not account for data that stays resident on the PIM side
across successive PIM regions, so a buffer whose producer and consumer are both on PIM
--- and which is therefore never moved at all --- is still charged every time. Second,
it treats every execution of a region as moving the full amount again: an input that
remains unchanged while an enclosing loop iterates is multiplied by the trip count all
the same.

\subsection{Performance Analysis}
\label{sec:micro}

\begin{figure*}[t]
  \centering
  \includegraphics[width=\textwidth]{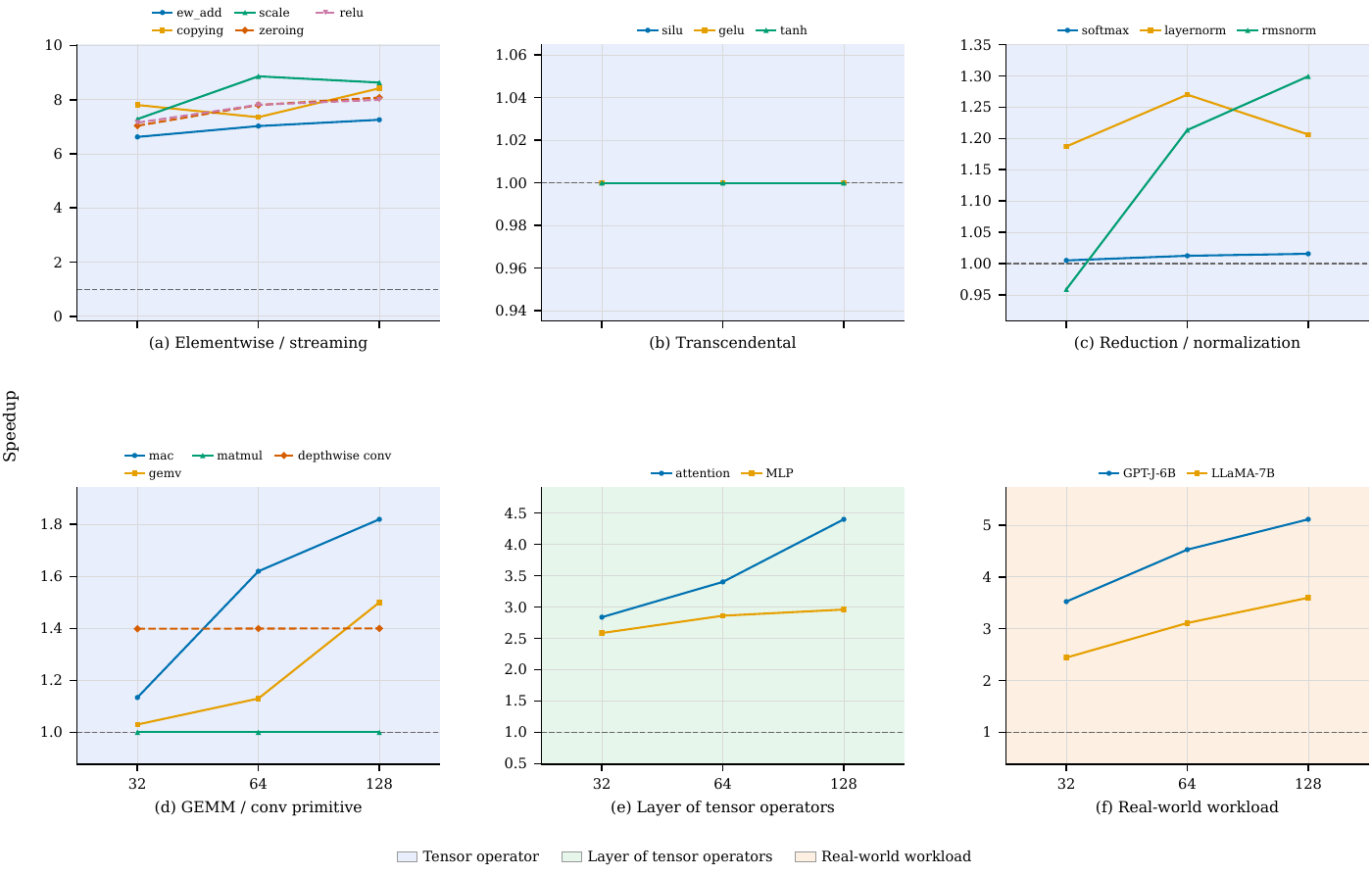}
  \caption{Speedup across core counts for the kernel benchmarks.}
  \label{fig:grid}
\end{figure*}

In the evaluation, the execution time of a workload, denoted as hybrid execution time $T_{hybrid}$, is computed by the following equation,
\begin{equation}
T_{hybrid}(N) = T_{host} + T_{pim}(N) + T_{copy} +N_{\mathrm{region}} T_{\mathrm{sw}},
\label{eq:decomp}
\end{equation}
where $T_{host}$ is the time spent in regions the rule left on the CPU,
$T_{pim}(N)$ is the execution time of the offloaded regions on $N$~PIM cores,
and $T_{copy}$ is the cost of moving their operands between host and PIM
memory explained in Section~\ref{sec:design:data_move_cost}. For $T_{\mathrm{sw}}$, we use the $2\,\mu s$ adopted by PIMProf~\cite{wei_pimproff}.

\paragraph{Elementwise/streaming.}
Every loop nest in this group is offloaded, leaving no host time outside the
offloaded regions. Speedup rises from 6.3$\times$ at 32 cores to \SI{7.3}{\times} at
128, a \SI{30}{\%} gain for an eightfold increase in cores, and the five kernels of
Fig.~\ref{fig:grid}(a) follow the same trajectory. The execution time is the sum of $T_{pim}$ and $T_{copy}$ as Torch-PIM offloaded the entire loop nests to PIM, and only $T_{pim}$
decreases with the core count. Operand movement accounts for \SI{7.1}{\%} of all-CPU
time and is independent of the core count, so each doubling of cores yields a
smaller gain than the last. The flattening of these curves comes from the cost
of host--PIM data movement, not from a limit on PIM throughput.

\paragraph{Transcendental.}
Fig.~\ref{fig:grid}(b) shows no improvements for workloads in \textit{Transcendental}.
No loop nest in this group is offloaded to PIM. They passed both C1 and C2, because the loop nests perform enough arithmetic per element. However, their
measured read bandwidth stays below $\theta$; they issue no non-temporal
stores, so they satisfy neither C3 nor C4. The whole benchmark therefore runs
on the host, and no speedup is observed.

\paragraph{Reduction/Normalization.}
Fig.~\ref{fig:grid}(c) depicts the performance of workloads in \textit{Reduction/normalization}.
\texttt{softmax} gains \SI{1.5}{\%} at 128 cores. One of its eight nests is offloaded;
the remaining seven, including the nest that evaluates the exponential, stay on
the host and account for \SI{97}{\%} of all-CPU time.

\texttt{layernorm} and \texttt{rmsnorm} capture \SI{48.5}{\%} and \SI{63.5}{\%} of all-CPU
time respectively, but behave differently with the core count.
\texttt{rmsnorm} improves monotonically, from \SI{0.96}{\times} at 32 cores to \SI{1.30}{\times} at 128, a gain of \SI{35}{\%}. At 32 cores, the operand movement cost $T_{copy}$ outweighs the PIM gain and the offloaded version runs \SI{4}{\%} slower than the host.
\texttt{layernorm} instead peaks at 64 cores, at
\SI{1.19}{\times}, \SI{1.27}{\times}, and \SI{1.21}{\times} for 32, 64 and 128 cores --- a \SI{6.7}{\%}
gain followed by a \SI{4.7}{\%} regression. The regression originates on the PIM
side: the PIM execution time of the offloaded regions rises by \SI{17.8}{\%} from 64
to 128 cores.

The difference lies in how much parallel work the offloaded regions contain.
Adding cores shrinks the per-core work but adds synchronization and result-aggregation overhead; the fewer parallel iterations a region has, the smaller the core count at which the overhead wins. \texttt{layernorm}'s offloaded regions are
smaller than \texttt{rmsnorm}'s, placing that crossover near 64 cores, whereas
\texttt{rmsnorm} does not reach it within 128.

\paragraph{GEMM/Convolution Primitives.}
Fig.~\ref{fig:grid}(d) depicts speed\-ups for the \textit{GEMM/convolution primitives}.
All four benchmarks in this group perform multiply-accumulate, yet the improvement
with offloading differs. What separates them is not the operation but the reuse per
element.

\texttt{matmul} is unchanged at \SI{1.00}{\times}. An $N \times N$ square GEMM
performs $2N^3$ FLOPs over $3N^2$ four-byte elements, an arithmetic intensity
of $N/6$ FLOP/byte, which is 341 at $N = 2{,}048$; reuse grows with $N$, so larger
instances are further from the memory-bound regime. The nest is the largest in
this group and passes Stage 1, but its reuse keeps tiles resident
in cache, reducing off-core read requests, and its measured bandwidth stays
below $\theta$. The rule thus establishes that the nest is large and parallel
enough, then rejects it at Stage 2 for not being memory-bound.

\texttt{gemv} performs the same multiply-accumulate at the opposite extreme.
Each matrix element is read once, and only the vector is reused, so memory
traffic per element is high. Its measured bandwidth exceeds $\theta$, and
the condition C3 holds, improving by 52\% from 32 to 128 cores to reach
\SI{1.56}{\times}. The same operation receives opposite verdicts on reuse alone, and
the rule arrives at the distinction by observing bandwidth rather than
computing reuse statically.

In \texttt{mac}, the nest that passes is not the multiply-accumulate loop but
the one preceding it. That loop fills a result array sequentially with no
reuse and uses non-temporal stores, which bypass the cache, so its write
traffic never appears in the off-core read counter. Its measured read bandwidth
is near zero, and it cannot pass C3. However, the fraction of backend cycles bound on
stores exceeds $\sigma$, and it enters through C4. \texttt{mac} is the only
benchmark that passes through C4 alone, and without the write-side path this
nest would appear in no counter at all. The benchmark improves by \SI{61}{\%}
from 32 to 128 cores, reaching \SI{1.82}{\times}.

For \texttt{depthwise\_conv} the speedup is flat at 1.40$\times$ across 32, 64
and 128 cores, even though the PIM execution time of the offloaded regions
falls in inverse proportion to the core count, halving over the same range. Of
the three components of hybrid execution time, only the PIM term varies with
the core count, and it already accounts for just 0.7\% of all-CPU time.
Halving it therefore shortens end-to-end time by 0.35\%, which moves the
speedup from 1.40$\times$ to roughly 1.405$\times$ --- below the second decimal
place. What dominates instead is the host time left outside the offloaded regions:
two of the three loop nests are offloaded, capturing 29\% of all-CPU time and
leaving 71\% on the host. This residual is independent of the core count and
bounds the attainable speedup at 1.408$\times$; the measured 1.40$\times$
reaches 99.4\% of that bound. The curve is flat not because PIM stops
improving, but because only 0.7\% of end-to-end time remains for any such
improvement to act on. The residual host time is the bulk convolution loop. Its reuse rate is high enough
that the measured read bandwidth stays below $\theta$, so it does not pass C3 , and it issues no non-temporal stores, so it does not enter through C4 either. This case is therefore a property of the workload rather than a failure of the
rule. Every memory-bound loop nest available for offloading was captured, and
the result attains 99.4\% of what that capture permits, so no further gain is
available on the decision side.

\paragraph{Layer of Tensor Operators and Real-World Workloads.}
Fig.~\ref{fig:grid}(e) and \ref{fig:grid}(f) show the improvements
for \textit{Layer of tensor operators} and \textit{Real-world workloads}.
\texttt{GPT-J-6B} and \texttt{LLaMA-7B} offload \SI{22}{\%} and \SI{24}{\%}
of their loop
nests, capturing \SI{97.0}{\%} and \SI{96.0}{\%} of all-CPU time. Speedup rises by \SI{45}{\%} from
32 to 128 cores for \texttt{GPT-J-6B}, reaching \SI{5.1}{\times}, while
\texttt{LLaMA-7B} reaches \SI{3.6}{\times}.

Every offloaded GEMM is a thin GEMM, that is, a GEMV. In an $M \times N \times
K$ product with small $M$, the weight matrix $[N \times K]$ is read once and
reuse exists only along $M$; the arithmetic intensity is $M/2$ FLOP/byte, or
0.5 at $M = 1$, which is memory-bound. This follows from the workload rather
than from the rule: LLM decoding generates one token at a time, so the batch
dimension is thin, and every projection becomes a weight-streaming operation.

\texttt{mlp} is the extreme case. At batch size one, every linear layer takes
the form $W \cdot x$; its two offloaded nests have a low reuse rate and
measured bandwidths of 6.97 and \SI{9.19}{\giga\byte/\s}, three to five times $\theta$. The
bandwidth is high because there is no reuse to exploit.

The offloaded sets of \texttt{GPT-J-6B} and \texttt{LLaMA-7B} show the same
structure in their measured signatures. One group has bandwidths of
\qtyrange[range-phrase = --,range-units = single]{15}{18}{\giga\byte/\s} with a store-bound fraction of zero --- nests 401, 901, 1201,
4001, 4901 and 5801 in \texttt{GPT-J-6B} and 301, 1701, 4501, 5401 and 6301 in
\texttt{LLaMA-7B} --- corresponding to the $W_q$, $W_k$, $W_v$, $W_o$ and FFN
projections, each streaming a $[d \times d]$ weight matrix once. The other has
bandwidths of \qtyrange[range-phrase = --,range-units = single]{3}{9}{\giga\byte/\s} with store-bound fractions of 0.07--0.15 and mixes
elementwise computation with stores.

The 50 and 51 loop nests in \texttt{GPT-J-6B} and \texttt{LLaMA-7B} rejected to be offloaded,
and 94\% of the rejections occur at Stage~1: most nests are eliminated on
size before memory boundedness is evaluated at all.

Host time outside the offloaded regions is \qty{3.0}{\percent} and \qty{4.0}{\percent}, so the remaining
gap lies in PIM execution and data movement rather than in regions the rule
failed to identify.

\subsection{Offload results can be different according to input sizes}

\begin{table}[t]
\centering
\small
\caption{Offloaded loop nests per operator as the problem size $N$ varies.
Each entry is $N\!\rightarrow\!k$, where $k$ of the operator's loop nests are
selected for offloading.}
\label{tab:size-sweep}
\begin{tabular}{llp{4.6cm}}
\toprule
Operator & Nests & Offloaded nests at each $N$ \\
\midrule
\multicolumn{3}{l}{\textit{Monotone}}\\
mac        & 3  & 256$\rightarrow$0, 1024$\rightarrow$0, 2048$\rightarrow$1, 4096$\rightarrow$2 \\
mlp        & 7  & 1024$\rightarrow$0, 4096$\rightarrow$2, 8192$\rightarrow$4 \\
attention  & 14 & 512/1024/2048$\rightarrow$0, 4096$\rightarrow$3, 8192$\rightarrow$3 \\
gemv       & 2  & 512/1024$\rightarrow$0, 4096/8192/16384$\rightarrow$1 \\
relu       & 1  & 512/1024$\rightarrow$0, $\geq$4096$\rightarrow$1 \\
ew\_add    & 1  & 512/1024$\rightarrow$0, $\geq$4096$\rightarrow$1 \\
\midrule
\multicolumn{3}{l}{\textit{Non-monotone}}\\
softmax    & 8  & 1024$\rightarrow$0, 4096$\rightarrow$2, 16384$\rightarrow$5, 32768$\rightarrow$4, 65536$\rightarrow$3 \\
\midrule
\multicolumn{3}{l}{\textit{Size-invariant}}\\
depthwise\_conv & 3  & 2 at all sizes \\
\bottomrule
\end{tabular}
\end{table}

Table~\ref{tab:size-sweep} reports the number of loop nests selected for
offloading across 11 operators and 54 loop nests as the problem size $N$
varies. Two observations follow.

First, the decision is made per loop nest, not per operator. In 7 of the 8
operators with more than one nest, only a subset of the nests is selected even
at the largest $N$: 3 of 14 for \texttt{attention}, 3 of 12 for \texttt{layernorm}, 3 of 8 for
\texttt{softmax}, 4 of 7 for \texttt{mlp}, 2 of 3 for \texttt{mac}, 2 of 3 for \texttt{depthwise convolution}, and
1 of 2 for \texttt{gemv}. A framework that treats an entire operator as a single
candidate cannot express this split.

Second, the selected nests depend on $N$. For \texttt{mac}, \texttt{mlp}, \texttt{attention}, and
\texttt{gemv}, no nest is selected at the smallest sizes, and nests cross the threshold
one at a time as $N$ grows. For \texttt{layernorm}, \texttt{depthwise convolution}, and \texttt{silu}, the
count is essentially fixed over the range measured.

\texttt{softmax} is not monotone. The count rises from 0 at $N = 1024$ to 2 at 4096 and
5 at 16384, then falls to 4 at 32768 and 3 at 65536. Its 8 nests flip at
different values of $N$, so a larger problem size does not uniformly favor
offloading.

\subsection{Decomposing the Performance Gap Across Granularities}
\label{sec:granularity-decomposition}
Table~\ref{tab:granularity-delta} compares the three granularities. The basic
block (BB) granularity and operator-level granularity lose for different
reasons: operator granularity raises $T_{\mathrm{pim}}$ from $29.5\%$ to
$38.8\%$ of the loop-granularity baseline, while BB granularity leaves
$T_{\mathrm{pim}}$ and $T_{\mathrm{cpu}}$ at their loop-granularity values and
raises $T_{\mathrm{copy}}$ from $2.1\%$ to $6.4\%$.

\begin{table}[t]
\caption{Cost of each granularity relative to loop granularity. All entries are
percentages of the end-to-end execution time under loop granularity.}
\label{tab:granularity-delta}
\small
\setlength{\tabcolsep}{8pt}
\begin{tabular}{@{}lrrrr@{}}
\toprule
& \multicolumn{2}{c}{operator $-$ loop}
& \multicolumn{2}{c}{BB $-$ loop} \\
\cmidrule(lr){2-3}\cmidrule(lr){4-5}
Benchmark & $\Delta T_{\mathrm{pim}}$ (\%) & Total (\%)
          & $\Delta T_{\mathrm{copy}}$ (\%) & Total (\%) \\
\midrule
GPT-J     & $+9.3$  & $+9.8$  & $+4.3$ & $+4.8$ \\
LLaMA     & $+6.2$  & $+6.5$  & $+3.2$ & $+3.5$ \\
attention & $+77.9$ & $+67.8$ & $+1.6$ & $+2.7$ \\
\bottomrule
\end{tabular}
\end{table}

Operator granularity places every loop nest of an operator on PIM as a unit, so
compute-bound loop nests inside an operator are offloaded along with the
memory-bound ones. On \texttt{GPT-J} this gives $\Delta T_{\mathrm{pim}} =$ \SI{+9.3}{\%},
$\Delta T_{\mathrm{cpu}} =$ \SI{-1.1}{\%} and $\Delta T_{\mathrm{copy}} =$ \SI{+1.6}{\%},
summing to \SI{+9.8}{\%}, of which \SI{95}{\%} is $\Delta T_{\mathrm{pim}}$: loop nests
that consume \SI{1.1}{\%} of the baseline on the CPU consume \SI{9.3}{\%} on PIM, a factor
of \SI{8.8}{\times}. Data movement accounts for \SI{16}{\%}, so the cost comes from the
placement decision rather than from moving data. On \texttt{attention} the same effect is
\SI{+77.9}{\%}, of which $10.1$ percentage points are recovered on the CPU and copy
sides, leaving \SI{+67.8}{\%}.

In the case of BB granularity, crossings rise from
$18$ to $372$, giving $\Delta T_{\mathrm{copy}} = +4.3\%$ and
$\Delta T_{\mathrm{switch}} = +0.4\%$ over $354$ additional crossings at
$2\,\mu\mathrm{s}$ each, for $+4.8\%$; $91\%$ of this is boundary tensor
materialization and $9\%$ the switching constant on \texttt{GPT-J}. The per-crossing copy
cost is $21.6\,\mu\mathrm{s}$ on \texttt{GPT-J} against $3.2\,\mu\mathrm{s}$ on attention,
scaling with boundary tensor size, so the fragmentation cost is the product of
crossing count and boundary tensor size, and it is this product that segment
merging keeps bounded.

\section{Limitations and Future Work}
\label{sec:limitations}

\paragraph{Empirical basis of the thresholds.}
$\theta$ and $\sigma$ are the only calibrated constants in the rule, and both
rest on a narrow base. We sweep $\theta$ on the same 44 loop nests over which we
report the $37/44 = 0.841$ agreement, and hold out no separate set, so that
figure states how well a single threshold separates this population rather than
how the calibrated rule transfers to nests it was not fitted on. Two false
positives persist at every $\theta \ge$ \SI{1.80}{\giga\byte/\s} and do not disappear up to
\SI{4.00}{\giga\byte/\s}, which indicates that measured read bandwidth alone does not separate
every region. $\sigma$ rests on a single observation: \texttt{mac} is the only
benchmark in which a nest issues non-temporal stores and therefore the only one
that enters through C4, so the write-side path is calibrated on one sample.
Establishing the portability of the ratio $\theta / B_\mathrm{cpu}$ across hosts
with different memory systems, and combining the two signals of C3 and C4 into a
single decision rather than a disjunction of two independently calibrated
thresholds, are both left to future work.

\paragraph{Direction of the movement cost.}
Equation~\ref{eq:tcopy} bounds the movement cost from above rather than
estimating it. A buffer that a producer writes and a consumer reads while both
remain on PIM is charged at every region boundary, and an operand that does not
change while an enclosing loop iterates is charged once per trip. The error is
one-sided, so the speedups we report are lower bounds on what the same placement
would achieve under an exact accounting. Narrowing the bound requires tracking
residency across consecutive PIM regions and identifying loop-invariant
operands, neither of which the current pipeline does. Torch-PIM emits offload
markers and leaves PIM-side code generation to the downstream toolchain, so the
PIM execution times we report reflect the quality of that toolchain as well as
the placement decision.

\paragraph{Scope of the candidate space.}
Torch-PIM requires every tensor extent to be a compile-time constant, which
holds for models exported at a fixed input shape but excludes models whose
shapes are resolved at run time. Within such a model, candidates are the loops
that bufferization materializes as \texttt{scf.parallel}. A loop that is
sequential as emitted but would become parallel under a transformation never
enters the candidate space, since the pipeline decides placement over the loops
it produces rather than over the loops it could produce. Broadening the
candidate space would require legality analysis for the transformations
concerned and a decision rule that ranks candidates the pipeline has not yet
materialized.

\section{Conclusion}
\label{sec:conclusion}

Torch-PIM lowers a PyTorch model through MLIR and decides host-versus-PIM
placement over the loop nests that progressive lowering materializes. Nothing
designates these candidates in advance: they are neither a list of operator
types, nor patterns identified beforehand, nor regions a programmer annotated.
Each candidate is judged in two stages, on the amount of work it carries and on
its memory boundedness, and every quantity the rule consumes is measured on the
host, so no candidate is ever executed or simulated on the PIM device. Our evaluations on PIM configurations of 32
to 128 cores reach speedups of up to
\qty{8.6}{\times} on tensor operators, \qty{2.9}{\times} on MLP,  \qty{4.4}{\times} on  Attention, \qty{5.1}{\times} on GPT-J-6B, and \qty{3.6}{\times} on decoder layers of LLaMA-7B  over CPU-only execution.

\bibliographystyle{ACM-Reference-Format}
\bibliography{citation}

\end{document}